# Deep Search for a Scattered Light Dust Halo Around Vega with the Hubble Space Telescope

Schuyler G. Wolff[1], András Gáspár[1], George H. Rieke[1], Jarron M. Leisenring[1], Kate Su[1], David J. Wilner[2], Luca Matrà[3], Marie Ygouf[4], and Nicholas P. Ballering[5]
[1] Steward Observatory and the Department of Astronomy, The University of Arizona, 933 N Cherry Avenue, Tucson, AZ 85721, USA; sgwolff@arizona.edu
[2] Center for Astrophysics, Harvard & Smithsonian, 60 Garden Street, Cambridge, MA 02138, USA
[3] School of Physics, Trinity College Dublin, the University of Dublin, College Green, Dublin 2, Ireland
[4] Jet Propulsion Laboratory, California Institute of Technology, Pasadena, CA 91109, USA
[5] Department of Astronomy, University of Virginia, 530 McCormick Road, Charlottesville, VA 22904, USA

## Abstract

We present a provisory scattered-light detection of the Vega debris disk using deep Hubble Space Telescope (HST) coronagraphy (PID 16666). At only 7.7 pc, Vega is immensely important in debris disk studies both for its prominence and also because it allows the highest physical resolution among all debris systems relative to temperature zones around the star. We employ the STIS coronagraph's widest wedge position and classical reference differential imaging to achieve among the lowest surface-brightness sensitivities to date ($\sim 4\ \mu$Jy arcsec$^{-2}$) at wide separations using 32 orbits in Cycle 29. We detect a halo extending from the inner edge of our effective inner working angle at 10″.5 out to the photon noise floor at 30″ (80–230 au). The face-on orientation of the system and the lack of a perfectly color-matched point-spread function star have posed significant challenges to the reductions, particularly regarding artifacts from the imperfect color matching. However, we find that a halo of small dust grains provides the best explanation for the observed signal. Unlike Fomalhaut (a close twin to Vega in luminosity, distance, and age), there is no clear distinction in scattered light between the parent planetesimal belt observed with the Atacama Large Millimeter/submillimeter Array and the extended dust halo. These HST observations complement JWST GTO Cycle 1 observations of the system with NIRCam and MIRI.

*Unified Astronomy Thesaurus concepts:* Debris disks (363); Circumstellar dust (236)

## 1. Introduction

The first debris disk was detected around Vega as an infrared excess when it was observed as an IRAS calibration source (Aumann et al. 1984). Termed "the Vega phenomenon," later surveys of IRAS data revealed unresolved dust emission around more than 100 main-sequence stars (Backman & Paresce 1993). Incredibly, this discovery of Vega's Kuiper Belt analog in 1984 preceded the finding the first (non-Pluto) Kuiper Belt object in 1992 (Jewitt & Luu 1993).

As one of the nearest (7.7 pc) and brightest (40 $L_\odot$) debris disk hosts, Vega has long been a proving ground for studies of debris disk physics and new observing techniques. The Vega system was among the first debris disks to be detected in the submillimeter with JCMT/SCUBA (Holland et al. 1998), with a radius of ∼85 au. The first very hot ($\lesssim 1500$ K) disk component within a radius of $\lesssim 4$ au was discovered around Vega through interferometry at 2 $\mu$m (Ciardi et al. 2001; Absil et al. 2006). Subsequent measurements place this dust within ∼0.2 au (Mennesson et al. 2011). Early Spitzer observations found a very extended dust halo surrounding Vega out to ∼400 au, providing evidence for a population of smaller, collisionally produced grains either escaping or at wide orbits (Su et al. 2005; Müller et al. 2010). As both the first discovered and one of the most accessible to detailed characterization, the Vega system is the archetype that forms the basis for comparison with other systems. For more details, see the reviews by Wyatt (2008) and Hughes et al. (2018).

In summary, the Vega debris system is complex, with (i) a hot component within 1 au (Absil et al. 2006; Mennesson et al. 2011); (ii) a warm component originally thought to lie at ∼14 au in analogy with our asteroid belt (Su et al. 2013), but shown to be more extended in new JWST/MIRI imaging (K. Su et al. 2024, in preparation); (iii) a cold debris ring (Kuiper Belt analog) with a peak at 11″ (85 au) resolved with Herschel at 70 $\mu$m (6″ beam; Sibthorpe et al. 2010) and the Atacama Large Millimeter/submillimeter Array (ALMA) at 1.3 mm (1″.8 × 0″.9 beam; Matrà et al. 2020; see also Marshall et al. 2022); and (iv) a halo of superheated grains extending out from the cold ring detected with Spitzer/MIPS at 24 $\mu$m (6″ beam; Su et al. 2005).

Vega has also been the target of many companion searches using both space- and ground-based facilities (e.g., Böhm et al. 2015; Meshkat et al. 2018). While there are no confirmed companion detections, radial velocity (RV) observations show a candidate companion with a period of 2.43 days, albeit with low confidence due to Vega's enhanced stellar activity (Hurt et al. 2021). Most recently, Ren et al. (2023) used the Keck vector vortex coronagraph to probe a region from 1 to 22 au and provide an upper limit on the companion mass of <3 $M_{\rm Jup}$ at 12 au. Recent Large Binocular Telescope Interferometer results (expanding upon Ertel et al. 2020) indicate the presence of substructure in the exozodi distribution, signaling the presence of a close-in, planetary-mass companion (V. Faramaz et al. 2024, in preparation) that may be related to the RV detections above. There are currently no known planetary-mass companions at a separation near the cold belt.

**Table 1**
HST/STIS Observations

| Date | Target | Visit | Aperture | Orientation (deg) | Total Integration Time (s) | No. of Frames |
|---|---|---|---|---|---|---|
| 2022-Apr-7 | Vega | 65 | WEDGEB2.8 | −127.4 | 310.0 | 31 |
| 2022-Apr-7 | Vega | 64 | WEDGEB2.8 | −132.4 | 310.0 | 31 |
| 2022-Apr-7 | α Cyg | 63 | WEDGEB2.8 | ⋯ | 979.2 | 31 |
| 2022-Apr-7 | Vega | 62 | WEDGEB2.8 | −147.4 | 310.0 | 31 |
| 2022-Apr-7 | Vega | 61 | WEDGEB2.8 | −164.4 | 310.0 | 31 |
| 2022-May-19 | Vega | 55 | WEDGEB2.8 | −161.9 | 310.0 | 31 |
| 2022-May-19 | Vega | 54 | WEDGEB2.8 | −172.9 | 310.0 | 31 |
| 2022-May-19 | α Cyg | 53 | WEDGEB2.8 | ⋯ | 1016.4 | 33 |
| 2022-May-19 | Vega | 52 | WEDGEB2.8 | 171.4 | 310.0 | 31 |
| 2022-May-19 | Vega | 51 | WEDGEB2.8 | 157.1 | 310.0 | 31 |
| 2022-Jun-24 | Vega | 45 | WEDGEB2.8 | 158.6 | 310.0 | 31 |
| 2022-Jun-24 | Vega | 44 | WEDGEB2.8 | 153.6 | 310.0 | 31 |
| 2022-Jun-24 | α Cyg | 43 | WEDGEB2.8 | ⋯ | 739.2 | 23 |
| 2022-Jun-24 | Vega | 42 | WEDGEB2.8 | 136.6 | 300 | 30 |
| 2022-Jun-24 | Vega | 41 | WEDGEB2.8 | 123.6 | 300 | 30 |
| 2022-Aug-13 | Vega | 35 | WEDGEB2.8 | 105.0 | 310.0 | 31 |
| 2022-Aug-13 | Vega | 34 | WEDGEB2.8 | 99.98 | 310.0 | 31 |
| 2022-Aug-13 | δ Cyg | 33 | WEDGEB2.8 | ⋯ | 1873.3 | 14 |
| 2022-Aug-13 | Vega | 32 | WEDGEB2.8 | 84.98 | 310.0 | 31 |
| 2022-Aug-13 | Vega | 31 | WEDGEB2.8 | 69.98 | 310.0 | 31 |
| 2022-Oct-4 | Vega | 24 | WEDGEB2.8 | 52.07 | 310.0 | 31 |
| 2022-Oct-4 | ζ Aql | 23 | WEDGEB2.8 | ⋯ | 1830 | 12 |
| 2022-Oct-4 | Vega | 22 | WEDGEB2.8 | 47.07 | 310.0 | 31 |
| 2022-Oct-4 | Vega | 21[*] | WEDGEB2.8 | 27.07 | 310.0 | 31 |
| 2023-Dec-16 | Vega | 15 | WEDGEB2.8 | −25.51 | 310.0 | 31 |
| 2023-Dec-16 | Vega | 14 | WEDGEB2.8 | −24.04 | 310.0 | 31 |
| 2023-Dec-16 | α Cyg | 13 | WEDGEB2.8 | ⋯ | 310.0 | 31 |
| 2023-Dec-16 | Vega | 12 | WEDGEB2.8 | −29.01 | 310.0 | 31 |
| 2023-Dec-16 | Vega | 11 | WEDGEB2.8 | −33.66 | 310.0 | 31 |
| 2023-Feb-21 | Vega | 4 | WEDGEB2.8 | −98.94 | 310.0 | 31 |
| 2023-Feb-21 | α Cyg | 3 | WEDGEB2.8 | ⋯ | 310.0 | 31 |
| 2023-Feb-21 | Vega | 2 | WEDGEB2.8 | −103.6 | 310.0 | 31 |
| 2023-Feb-21 | Vega | 1 | WEDGEB2.8 | −107.9 | 310.0 | 31 |

**Note.** Visit 21 failed due to an unsuccessful guide star acquisition and was repeated as visit 15.

We focus here on the outer-disk components accessible with Hubble Space Telescope (HST) coronagraphy, namely, the Kuiper Belt analog and the dust halo. The highest-resolution images of the outer ring were obtained with ALMA by Matrà et al. (2020). The data show a prominent, broad ring with a peak in surface brightness at 11″ (85 au), a well-defined inner edge, and extending out to at least 150 au, although further extent at low surface brightness may have been suppressed due to the lack of short $u-v$ spacings. Modeling of the ALMA data suggests that the inner edge of the ring is likely sharp, although a softer-edge Gaussian fit is also possible. Copious dust is produced in this ring, as shown by the bright halo emanating from it and detected at 24 $\mu$m (Su et al. 2006). Thus, its structure will be highlighted by light scattered off these grains and, like the outer ring around the very similar star Fomalhaut, might be apparent in coronagraphic images.

While ground-based adaptive optics systems have been highly efficient at resolving higher-surface-brightness and compact disks, the enormous spatial extent (∼25″) and low surface brightness of the Vega disk place it well outside of their reach due to background and field-of-view (FOV) limits. Furthermore, the face-on orientation of the disk makes many point-spread function (PSF) subtraction techniques that rely on azimuthal variation unsuitable, and misses the forward scattering peak at small scattering angles. The STIS coronagraph on the HST provides unprecedented stability to PSF variations and is ideally suited to detect faint disk features at all spatial angles. This paper describes HST/STIS observations of the Vega system that may reveal the first scattered-light detection. This conclusion is subject to our overcoming significant challenges in the data reduction, as discussed in the Appendices.

## 2. HST/STIS Observations

We used HST/STIS to obtain deep coronagraphic observations of Vega and a series of PSF reference stars in the Cycle 29 GO program 16666 (PI: Wolff). Details of the observations are provided in Table 1. The Vega disk is uniquely challenging to observe: The disk is viewed face-on (so scattering will be inefficient; Hedman & Stark 2015), and (from the Herschel and ALMA images; Sibthorpe et al. 2010; Matrà et al. 2020) radially extended with a low surface brightness and requiring a large FOV instrument to observe. To achieve the deepest possible contrast, we elected to use the widest wedge position available (to block the maximum amount of stellar light), the WB2.8 location of the 50CORON coronagraphic mask of the 50CCD detector, with minimal but intentional saturation of the inner core. These observations include contemporaneous, interleaved, and identically observed (same total counts/pixel/exposure at small separations) imaging of color-matched, PSF template stars.

A total of 32 orbits were split into seven visit groups, each consisting of four or five orbits. The third orbit in each group was reserved for the PSF reference while the remaining orbits observed Vega, resulting in a total of 25 science orbits and seven PSF orbits. Data were obtained beginning in 2022 April and continued through 2023 February, which allowed for full azimuthal coverage of the disk with changing celestial orientation angles (see Table 1). The observations within each set were executed in sequential, contiguous orbits (i.e., back to back with no interruptions except Earth occultation), to ensure PSF stability. The multi-roll technique we used has been proven in multiple HST programs (e.g., GO program 12228) to produce the optimal results for low-surface-brightness extended emission (Schneider et al. 2014), as the high number of roll dithers reduces systematic errors and image artifacts while also providing a full 360° imaging coverage of the target region.

The Vega observations had an exposure time of 10 s with 31 frames per orbit, resulting in a total integration time per orbit of 310 s and a total exposure time of 7750 s for the entire Vega program. This exposure time was selected based on HST GO program 13726 wherein Vega was used as a PSF calibration source using a slightly smaller wedge position of WEDGEB2.5. In that case, 10 s exposures led to some saturation but resulted in a useful inner working angle of ∼4″, which is adequate for our science goals of spatially resolving the disk morphology for Vega's Kuiper Belt analog. Our choice of the wider WEDGEB2.8 results in less saturation (though some saturation remains inside of ∼4″).

### 3. Choice of PSF Reference

Selecting the optimal PSF star for Vega proved challenging. Ideally, the PSF template star should be (i) relatively close in the sky to Vega, minimizing "breathing" variations in the PSF due to thermal instability of the telescope optical tube assembly (OTA); (ii) a close color match to Vega given the wide, unfiltered bandpass of the STIS coronagraph; and (iii) sufficiently bright as to provide similar signal-to-noise ratio (S/N) to Vega at wide separations and to reproduce detector behavior.

The only nearby, sufficiently bright source with a similar spectral type is $\alpha$ Cyg (Deneb, spectral type = A2Ia, $V_{\rm mag}$ = 1.25). This was the primary PSF template used in the program, but was shown to be a less ideal color match than initially assumed. Figure 1 compares the spectra convolved with the STIS bandpass for our selected PSF targets. $\alpha$ Cyg is intrinsically similar to Vega, but it is redder with $\Delta B - V = 0.09$.

It has been shown in Grady et al. (2003) that the broad bandpass and relatively diffuse PSF of STIS creates a color-dependent PSF shape, resulting in subtraction artifacts most prominent inside of 2″ but extending out to ∼10″. This effect can be modeled using the Tiny Tim software to simulate the STIS PSF for different stellar colors. However, Tiny Tim only simulates the PSF out to a separation of 4″.5. Unfortunately, a good estimate of the theoretical PSF outside of this region is not possible since the mid- and high-spatial-frequency errors in the optics are not well characterized, and those create the scattered light that dominates at large separations. Given that we purposefully saturate out to a separation of ∼4″, it was unclear how the color mismatch would present at wider separations.

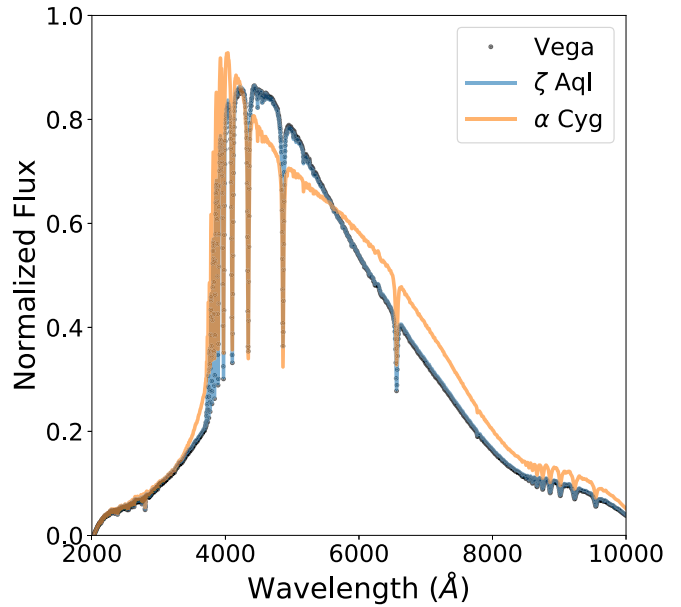

**Figure 1.** Spectral comparison between Vega and the two PSF calibration sources: $\alpha$ Cyg and $\zeta$ Aql. The spectra have been convolved with the STIS 50CCD transmission function available through the `pysynphot` package (i.e., the unfiltered sensitivity or detector quantum efficiency). While all targets are close spectral types, the fainter $\zeta$ Aql is a much better color match to Vega than $\alpha$ Cyg. The spectra are based on the BOSZ theoretical library (Bohlin et al. 2017) and the extinction law from Gordon et al. (2021), plus the measured stellar parameters, e.g., $T_{\rm eff}$, log($g$), and $A_V$.

Early PSF subtractions of $\alpha$ Cyg from Vega showed a radial profile shape characteristic of a color mismatch (see Figures 3 and 4 of Grady et al. 2003), albeit at wider separations with negative values close to the inner working angle (<5″) and a bright ring at ∼7″. To determine if these close-in features were related to a disk structure or solely from color-mismatch PSF artifacts, we conducted a search for a perfect color match to use as a PSF calibrator. Lamentably, the only nearby sources that were sufficiently bright and a near perfect color match were known to host companions. In the fourth visit group, we replaced $\alpha$ Cyg with the slightly fainter $\delta$ Cyg (A0IV, $V_{\rm mag}$ = 2.87), but the stellar companions proved too difficult to subtract out using an empirical PSF. We make no further reference to this PSF target. In the fifth visit group, we used the fainter $\zeta$ Aql (A0IV, $V_{\rm mag}$ = 2.99) as the PSF calibrator, which is a very close match to the colors of Vega (Johnson & Mitchell 1975), as shown in Figure 1. Here we were able to successfully mask the companion ($V_{\rm mag}$ = 12 at a separation of 7″.2). This limited our azimuthal coverage but provided a PSF reference with sufficient S/N inside of 10″ to test the color dependence of the PSF.

Exposure times for the PSF calibrators were chosen to reproduce the saturation pattern and radial brightness profile of the Vega observations. We scaled the Vega exposure time of 10 s based on the $V$ magnitudes to arrive at individual exposure times of 30.8 s for $\alpha$ Cyg and 152.5 s for $\zeta$ Aql. $\alpha$ Cyg was observed in five epochs with 31 integrations per epoch, resulting in a total exposure time of 4774 s. $\zeta$ Aql was only observed for a single epoch with 12 integrations for a total exposure time of 1830 s.

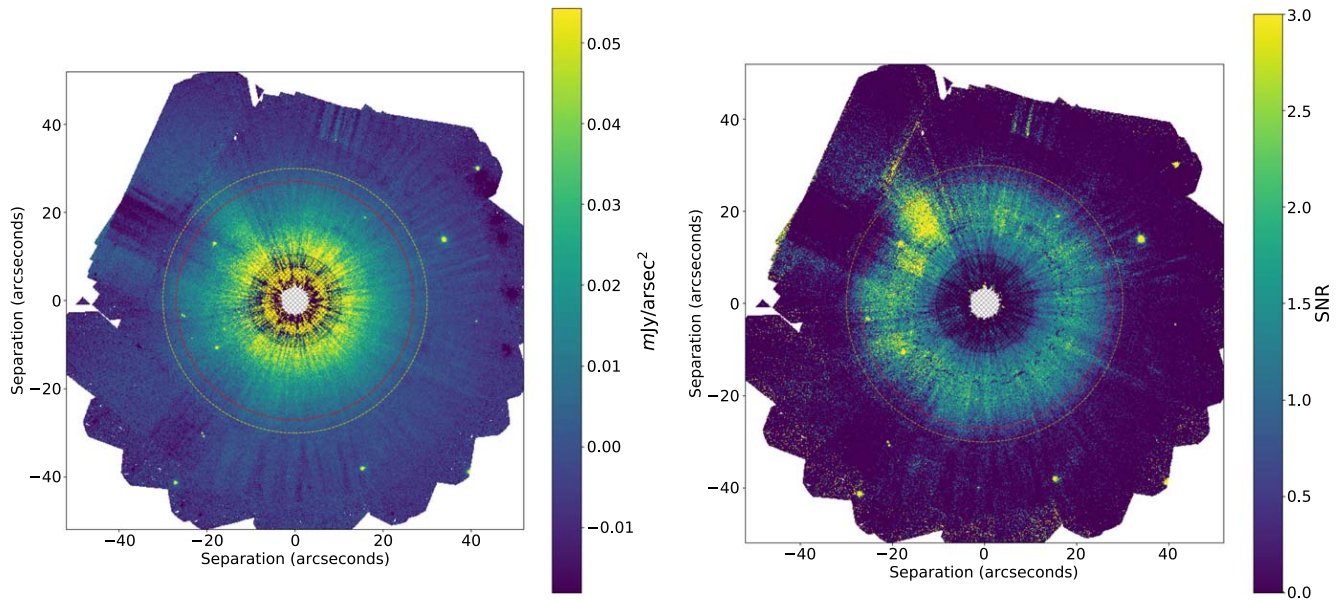

**Figure 2.** PSF subtraction results for the Vega system. Left: the scattered-light image of Vega using α Cyg as a PSF template presented in millijanskys per square arcsecond. Right: the signal-to-noise ratio (S/N) of the scattered-light disk detection using the uncertainty map described in Section 4 and presented in the Appendix. Both images are presented north up, east left. While the signal interior to 10.″5 (cross-hatched region) is indistinguishable from PSF residuals due to a color mismatch, the signal from 10.″5 to 27″ has an S/N > 1 per pixel and is consistent with a face-on disk. The scattered-light surface brightness drops off quickly between 27″ (red dashed line) and 30″ (yellow dashed line).

## 4. Data Reduction and Reference PSF Subtraction

Image calibration for each individual exposure was performed using the methods detailed in Wolff et al. (2023). First, we removed the video noise pattern (also known as the "Herringbone pattern" for STIS) using the Autofillet package (Jansen et al. 2003). Next the video-noise-corrected coronagraphic images were re-reduced following the standard image reduction steps using the calstis pipeline, available within the stistools package (part of Astroconda), maintained by the Space Telescope Science Institute (STScI). The result was a total of 775 individual science exposures and 167 reference exposures in units of counts per second. Masks for each exposure were generated to include STIS wedge positions, the OTA diffraction spikes, hot pixels and corresponding bleeding lines, saturated regions, any bright point sources (and in the case of ζ Aql a region surrounding the stellar companion), and any apparent imaging artifacts found in a by-eye examination. Finally, for each orbit (i.e., unique orientation angle) the images and masks were median combined.

For low-surface-brightness disk emission, classical reference differential imaging (cRDI; PSF subtraction with a dedicated reference star) outperforms other principal-component-analysis-based PSF subtraction methods (e.g., Ren et al. 2018; Wolff et al. 2023) for HST/STIS, where the PSF of the telescope OTA is relatively stable. Here we present the results of cRDI using α Cyg as the PSF reference star to maximize S/N at wide separations. We employ ζ Aql, the other observed PSF reference star, as a test of the color dependence in Section 4.1.

We find that the order of PSF subtraction steps has an impact on the final image sensitivity, and seek to limit numerical/processing noise by first combining the frames in an exposure taken over a single orbit with a unique telescope orientation angle. Each resulting image was background subtracted using the median value of a large region near the far edge of the image (with a separation of ∼45″) from the star. The background subtraction levels were amended after subtraction to result in a radial profile with a photon noise floor mean of zero counts s$^{-1}$. The location of the central star behind the coronagraphic mask for each frame was determined using a Radon transform algorithm to find the intersection point of the OTA diffraction spikes with a demonstrated precision of 0.1 pixels (Ren et al. 2019a). This is discussed in more detail in Appendix A. Using these centroids, the science images (and masks) were recentered and rotated to north up, while the reference star images (and masks) were recentered and rotated by the same angle to coalign the detector axes. Following translation, the mask values below 1 were rounded down to zero, thereby conservatively expanding the size of the mask by at most 1 pixel.

A scale factor between the science and reference images was determined by comparing mean counts. A ratio of the mean counts per integration in the science versus reference orbit stacks gave a value of 2.99 for Vega/α Cyg. This is within 3% of the nominal value of 3.08 based on the $V$ magnitudes. However, α Cyg is known to be a variable star with a period of 11.7 days (Abt 1957), and we obtain a cleaner subtraction if we allow the scale factor to vary between the five orbits. We find the best Vega/α Cyg scale factors to be 2.98, 2.94, 2.96, 2.96, and 3.00 for orbits 03, 13, 43, 53, and 63, respectively (see Table 1). This aligns with the ∼5% variation expected due to stellar variability and telescope breathing. Finally, the reference image taken closest in time to each science image was determined and a scaled subtraction was performed.

The final subtracted image stacks (in units of counts per second) are median combined. The PHOTFLAM header keyword and the STIS pixel scale were used to convert to physical units of millijanskys per square arcsecond. The final cRDI-subtracted image is shown in Figure 2. The image shows signal consistent with a face-on scattered-light disk and several point sources. These point sources will be checked for common proper motion when combined with JWST observations (PID 1193) using NIRCam coronagraphy with planet mass sensitivities down to 0.75 $M_{Jup}$ at 4″ based on NIRCam's on-sky detector and coronagraphic performance (Carter et al. 2023; Rieke et al. 2023). There is a slight surface-brightness enhancement to the northeast, however this corresponds to the location of the lowest orbital coverage in the image and is unlikely to be significant.

The local error in each exposure will be dominated by position-dependent photon noise, which is overwhelmingly from the central star (most of which is subtracted) rather than the disk. However, in the PSF-subtracted images, the detector noise and disk-associated photon noise are also significant, with additional uncertainty contributions from the PSF centering error, telescope breathing, and focus changes across all 32 orbits. The propagated uncertainty map is estimated using the method described in Ren et al. (2019b). In short, a standard deviation map of the PSF-subtracted image stack (oriented to detector coordinates) is duplicated to match the number of unique science orientations and rotated north up before being combined in quadrature. The method is described in more detail in Appendix A.1 with the uncertainty map given in Figure 12. The corresponding S/N map is shown in the right panel of Figure 2. The scattered-light disk is detected with an S/N > 1 per pixel between 10″.5 and 27″, with a steep falloff of the surface brightness out to 30″ where the photon noise floor is reached. Integrated over the $R_{80} = 0″.16$ aperture (radius in arcseconds of an aperture that encloses 80% of the flux of a point source for the STIS 50CORON mode), this gives a mean S/N of ~40 per resolution element in the disk region. Interior to 10″.5, the noise is too high for the disk to be detected reliably.

### 4.1. Color-dependent Uncertainties

The large image uncertainties interior to 10″.5 are likely dominated by PSF residuals related to the slight color mismatch between the science target, Vega, and our PSF calibration source $\alpha$ Cyg. We investigate the color dependence of the cRDI PSF subtraction method in this section. In addition to the Vega − $\alpha$ Cyg reduction, we also performed a PSF subtraction using $\zeta$ Aql as the science target to reproduce any color-induced PSF artifacts. The $\zeta$ Aql data set contains only a single orbit and is significantly lower S/N than the Vega observations. To counteract this, we generate copies of the $\zeta$ Aql frames at different orientations (designed to match the position angle, PA, coverage of the 25 Vega observations) to smooth out any background sources and image artifacts when median combined. These are then used as science frames and subtracted using the $\alpha$ Cyg PSF identically to the reductions described above, using the same orbit-dependent scale factors. Given the close color match between Vega and $\zeta$ Aql, if the signal persists in this reduction it is due to a color mismatch with $\alpha$ Cyg rather than signal from disk scattered light.

As expected, the $\zeta$ Aql − $\alpha$ Cyg reduction shows large discrepancies from the Vega − $\alpha$ Cyg reduction inside of 10″, highlighting the impacts of the color mismatch, and we are unable to constrain any disk morphology in this region (see Figure 3). Outside of this region, there is no indication of a signal nearly as bright as the halo detected around Vega.

Azimuthally averaged radial profiles were generated (Figure 3), with 5$\sigma$ uncertainties provided by the uncertainty maps presented in Figures 12 and 13. For each radial bin, the uncertainties were combined using $\sigma = \sqrt{\frac{\sum \sigma_i^2}{N}}$, assuming uncorrelated noise. While the $\zeta$ Aql reduction is lower in flux than the Vega reduction (due to the fainter magnitude of the central star), the comparative shapes of the radial profiles elucidate any impacts of a color mismatch between the science targets (Vega and $\zeta$ Aql) and the PSF reference ($\alpha$ Cyg). Inside of ~8″, the Vega − $\alpha$ Cyg subtraction shows a bright ring with a peak at 5″, while the $\zeta$ Aql − $\alpha$ Cyg reduction shows a negative profile inside of 5″ and a sharply sloped increase to a peak at ~9″. This could indicate a bright inner disk surrounding Vega, in agreement with JWST/MIRI imaging observations (K. Su et al. 2024, in preparation), but it is not detected at sufficient S/N to be conclusive.

Outside of our effective inner working angle of 10″.5, the $\zeta$ Aql − $\alpha$ Cyg reduction is much shallower than Vega − $\alpha$ Cyg. We measure the slope of the radial profile between 10″.5 and 27″ to be −2.3 for Vega and −0.5 for $\zeta$ Aql. This implies that the color of the central star may be playing a role at these wide separations, though residual companion flux may also play a role here. Comparisons to the direction and magnitude of the color mismatch explored in Grady et al. (2003) agree with the $\zeta$ Aql − $\alpha$ Cyg radial profile; a subtraction normalized to the wings of the PSF will result in an oversubtraction in the innermost regions of the bluer $\zeta$ Aql and a slight undersubtraction in the outer regions. Furthermore, the magnitude of the undersubtraction appears nearly constant in radius, regardless of the $\Delta B - V$ color (see Figure 3 of Grady et al. 2003). Consequently, the sharp drop-off of flux in the outer regions of the Vega − $\alpha$ Cyg image is unlikely to result from the $\Delta B - V = 0.09$ difference. A color mismatch between the science and reference PSF is not able to completely explain the scattered-light signal observed around Vega. Outside of 27″, the Vega subtraction shows a steep decline out to ~30″ where the photon noise floor is reached. For a more detailed discussion of science and reference PSF color-induced "coronas," see Appendix A.3.

Ultimately, the precise disk morphology is ambiguous given these observations, though it is indicated that excess emission is present in a spatially extended halo, possibly from scattering by small dust grains. Future dedicated observations of a higher-S/N color-matched PSF calibrator may help to overcome this uncertainty but would require use of a noncontemporaneous observation, and thus be dependent on the thermal OTA response. More detailed modeling of the color-dependent impact of breathing and focus changes on the theoretical PSF at wide separations would also help but is unlikely to be completely satisfactory.

## 5. Analysis and Discussion

We have presented evidence for a scattered-light halo of dust particles surrounding Vega from 10″.5–27″ (80–210 au). We explore the possibility that the observed halo is a PSF-subtraction artifact extensively in Appendix A. First, the signal persists across all 25 Vega orbits, and is thus unlikely to be a PSF artifact related to breathing, a change in focus, or PSF centering uncertainties. Second, the observed radial surface-brightness distribution is not consistent with the difference in color between the science and reference targets. Finally, the steep drop from 27″–30″ persists regardless of the choice of scale factors between the science and reference images, and is inconsistent with charge transfer efficiency (CTE) effects. While the unique observing scenario of this data set (saturation pattern, wide wedge position, atypically low surface-brightness sensitivity) could result in unprecedented PSF behavior at wide separations, we conclude that light scattered off of small dust grains is the most likely source of the observed signal.

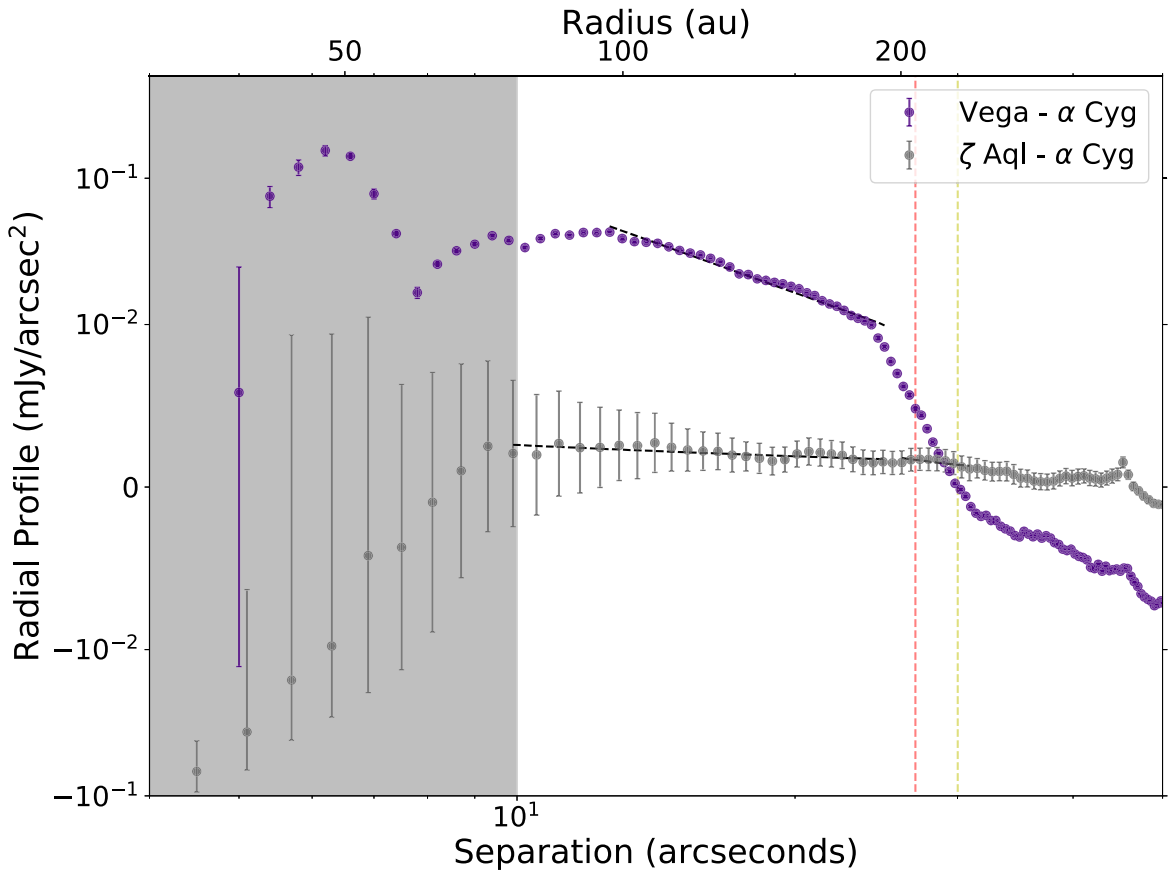

**Figure 3.** Radial profiles of the Vega PSF-subtracted image (Figure 2) are presented. For comparison, we also present the radial profile from the $\zeta$ Aql $-\alpha$ Cyg subtraction (gray) to test the impacts of the color-dependent PSF. Uncertainties are generated using the radial profile of the respective uncertainty maps. The Vega reductions are binned to an 8 pixel radius, while the noisier $\zeta$ Aql $-\alpha$ Cyg reduction uses 12 pixel bins. The y-axis is shown on a logarithmic scale. The profile shapes differ significantly inside of $\sim 8''$, and the data in this region are untrustworthy. Outside of this region, the Vega $-\alpha$ Cyg subtraction shows a halo of disk material extending out to at least $\sim 27''$ (210 au). The difference in the S/N achieved for $\alpha$ Cyg and $\zeta$ Aql is evident in the uncertainties.

In the following section, we investigate the dust properties required to produce the observed scattered-light levels. The large effective inner working angle and the 1% uncertainty in the scale factor between the science and reference images (see Appendix A.2) complicates the interpretation of the total scattered-light contribution around Vega. We provide an estimate of the scattering optical depth and a lower limit for the 90° scattering albedo to test the plausibility that small dust grains are responsible for the observed signal. Finally, it is shown that a simple dust model for the disk is able to reproduce the observed scattered-light flux levels without overproducing flux at longer wavelengths.

We estimate the radially dependent scattering optical depth, $\tau_{\text{scatt}}(r)$, over the region of interest using the scattered-light surface-brightness radial profile $I(r)$ given in Figure 3 via $\tau_{\text{scatt}}(r) = \Omega_r \times I(r)/F_{\nu,*}$ where $\Omega_r = \pi \frac{r^2}{d^2}$ for distance $d = 7.7$ pc and the flux density of Vega in the STIS bandpass $F_{\nu,*} = 2.9 \times 10^3$ Jy. The scattering optical depth is relatively flat with a mean value of $6.7^{+3.2}_{-1.7} \times 10^{-6}$ from $10.''5-27''$, with the uncertainties derived from the 1% variation in the Vega/$\alpha$ Cyg scale factor. This value is consistent with the nondetection of the disk with JWST/NIRCam with a predicted sensitivity limit of $\tau_{\text{scatt}}(r) < 2 - 8 \times 10^{-5}$ (C. Beichman et al. 2024, in preparation). As expected, it is also below the optical depth for the thermal emission derived from the JWST/MIRI color temperatures $\tau_{\text{em}} \sim 2 \times 10^{-5}$ (K. Su et al. 2024, in preparation).

A lower limit on the 90° dust scattering albedo can be estimated by comparing the scattered light and infrared fractional luminosities via $Q = \frac{f_{\text{Lscatt}}}{f_{\text{Lscatt}} + f_{\text{LIR}}}$. To estimate the total flux contributions from the Vega debris disk in scattered light, we sum the flux contained in the Vega $-\alpha$ Cyg reduction over the region of interest ($10.''5-30''$) to arrive at $f_{\text{Lscatt}} = 1.4 \times 10^{-5}$.[6] Using a value of $f_{\text{LIR}} = 3.1 \times 10^{-5}$ (Cotten & Song 2016), we estimate a 90° scattering angle albedo of $\gtrsim 0.31$. Note that this is a lower limit since it does not include scattered-light contributions from the disk inside of $10.''5$, about 3% of the disk surface area.[7] The uncertainty in the Vega/$\alpha$ Cyg scale factor accounts for an additional $\pm 0.1$ of the albedo uncertainty.

This albedo is in line with other measured debris disk albedos (see Figure 6, Table A.2 in Ren et al. 2023) though on the higher end, based on the 80°–100° scattering angles, and is intermediate to HD 191089 (0.23) and TWA 7 (0.57). Coupled with the radially extended scattered-light detection in this face-on disk, this implies a large population of small grains. Dust particles smaller than the observing wavelength scatter more isotropically, and the strong forward scattering peak observed in several debris disks systems (e.g., HR 4796A; Milli et al. 2017) is mitigated. Vega is very luminous, with a blowout size of $\sim 5$ $\mu$m predicted for astronomical silicates by setting the ratio of radiation pressure and gravitational force $\beta = 0.5$ (see Müller et al. 2010). Composition and porosity are also key players in determining the blowout size. Arnold et al. (2019) compares predicted blowout sizes for more complex aggregates and a range of compositions with Mie spheres in several nearby systems. For HR 4796A, similar in mass, effective temperature, and luminosity to Vega, they find blowout sizes of 2 to $>10$ $\mu$m depending on the porosity and composition. The strong Vega scattered-light detection at $\sim 0.6$ $\mu$m suggests a large population of grains small enough to be blown out of the system on hyperbolic trajectories. We further discuss the source of this dust population in Section 5.2.

---
[6] The final integrated Vega debris disk scattered-light flux is $8.5 \times 10^4$ counts s$^{-1}$. The stsynphot package is used to compute the total flux contributions for Vega using the STIS 50CCD a2d4 mode with $L_* = 6.1 \times 10^9$ counts s$^{-1}$.
[7] The sharp decline in the scattered-light radial profile outside of $27''$ (210 au) indicates a lack of significant scattered-light surface-brightness contributions outside of this radius.

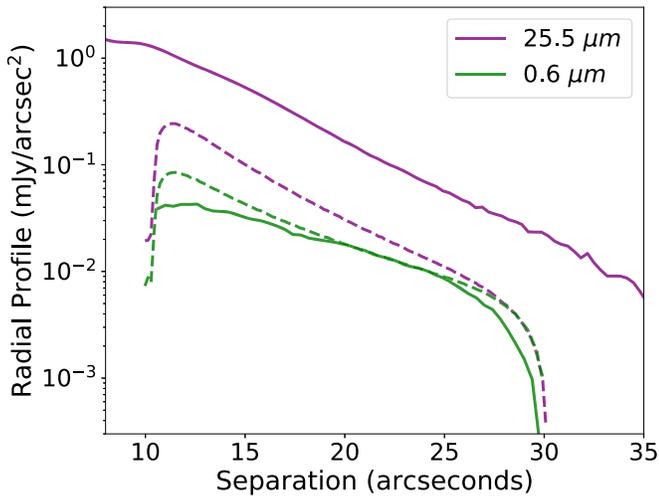

**Figure 4.** Here we compare radial profiles of the scattered light observed with HST/STIS and the thermal emission observed with JWST/MIRI to a simple dust model. The data are shown with solid lines and the models with dashed lines. The model includes a dust population comprised of water ice with a size distribution from 0.2 to 5 $\mu$m, a power-law slope of $-3.65$, a radial dust density distribution $\propto r^{-1}$, and a radius from $10''$ to $30''$. The model is able to reproduce the observed scattered-light surface-brightness levels without overproducing the flux observed in thermal emission.

As a final test of the credibility of the scattered-light detection, we generate a simple dust model and determine if it can reproduce the observed scattered-light surface brightness without overproducing the observed thermal emission distribution. Using the DiskDyn dust modeling code (Gáspár 2020), the model assumes a dust population comprised of water-ice grains with a size distribution extending from 0.2 to 5 $\mu$m with a power-law slope of $-3.65$. The disk extends from $10''$ to $30''$ with a radial dust distribution $n(r) \propto r^{-1}$, designed to roughly match the shape of the scattered-light surface density profile.

Model disk images are generated at the central wavelength of the STIS coronagraph (0.6 $\mu$m) and at 25.5 $\mu$m to match the JWST F2550W filter. Figure 4 compares the radial profiles of these models to the STIS profile presented in Figure 3 and the JWST/MIRI profile presented in Figure 5 (K. Su et al. 2024, in preparation). In this case, signal from the dust population required to reproduce the scattered light falls well below the observed thermal emission of the dust as observed with JWST. The paucity of thermal emission in the model may result either from a different grain population or from impurities in the hypothesized icy grains that increase their emissivity. While a complete exploration of the dust properties is beyond the scope of this work, it is clear that scenarios exist which are capable of producing this degree of scattered light while remaining consistent with the thermal emission information available in the literature.

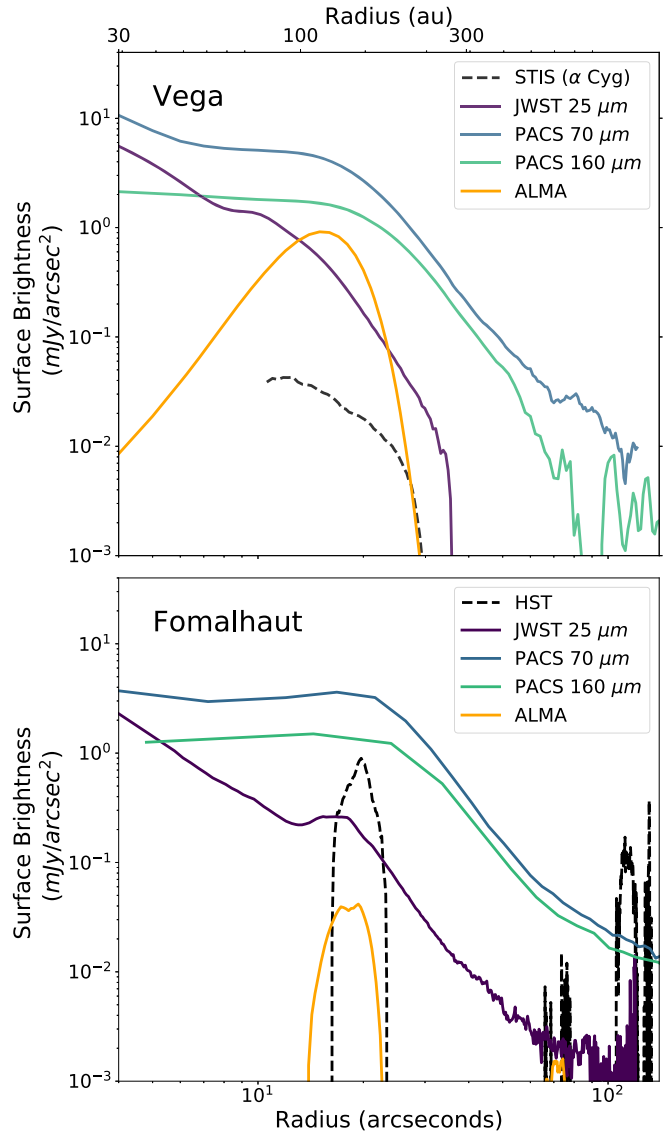

**Figure 5.** Comparison of the wavelength-dependent radial profiles for the Vega (top) and Fomalhaut (bottom) debris disks. The radial profiles for Vega are provided for HST/STIS (this work), JWST/MIRI (K. Su et al. 2024, in preparation), Herschel/PACS (Su et al. 2013), and ALMA (using the Gaussian model from Matrà et al. 2020). Radial profiles for Fomalhaut are adapted from HST/STIS (Gáspár et al. 2020), JWST/MIRI (Gáspár et al. 2023), Herschel/PACS (Su et al. 2013), and ALMA (MacGregor et al. 2017).

We compare radial profiles for these scattered-light observations to those of thermal emission available in the literature in Section 5.1 and discuss the source of the scattered light in more detail in Section 5.2. Section 5.3 considers Vega in context with the population of observed scattered-light halos.

### 5.1. Comparison to Fomalhaut

Vega allows the highest physical resolution relative to thermal zones around any star with an infrared excess—that is, relative to positions matched in stellar irradiance; for example, at the relative luminosities of Vega ($40\,L_\odot$) and Fomalhaut ($16\,L_\odot$), both at 7.7 pc, the equivalent thermal zones are at $\sqrt{40/16} \approx 1.6$ larger radii around Vega than Fomalhaut. Fomalhaut is an apt cognate for Vega, as one of the few other stars with a fractional disk luminosity $f < 10^{-4}$ for which a debris disk has been resolved in scattered light (see also the inconclusive result for HR 8799; Gerard et al. 2016). For comparison, multiwavelength radial profiles are presented for Vega and Fomalhaut in Figure 5. While the Herschel images show a similar radial profile for both systems (partly due to the larger beam size), they show very different profiles both at shorter wavelengths in scattered light and at longer wavelengths that trace the parent planetesimal populations.

Vega's fellow archetypical disk, Fomalhaut, has been extensively studied in scattered light (Kalas et al. 2005; Gáspár et al. 2020) and shows a narrow cold belt from 130 to 150 au (possibly confined by an undetected planet; e.g., Boley et al. 2012) with a scattered-light halo extending outwards from the belt. The parent planetesimal belt is clearly detected with ALMA (MacGregor et al. 2017), with an estimated dust mass of $0.015 \pm 0.010\,M_\oplus$. Fomalhaut has also recently been the target of groundbreaking JWST observations (Gáspár et al. 2023) showing the same narrow cold belt (albeit with a slight radial offset indicative of grain size stratification) and halo.

Despite the similarity in ages, distances, and large stellar luminosities, the Vega disk morphology differs significantly from Fomalhaut, with a much broader outer belt (see Figure 5). The shape of the radial profile is largely independent of wavelength, showing no clear stratification of dust particle sizes. A sharp, distinct cold belt is not observed around Vega in scattered light. Vega's outer belt is broader in extent than the Fomalhaut counterpart for both the small dust population (probed by HST) and the parent planetesimal population (as seen by ALMA). This implies for the Vega system that there is no analog to the Fomalhaut planet responsible for shepherding the dust into a narrow ring (over a wide distribution in dust particle sizes). We consider two scenarios to describe the scattered-light dust distribution: (i) the outer belt contains a population of scattered planetesimals driven by a massive planet, resulting in the collisional replenishment of the smaller dust particles; or (ii) Vega's outer belt is dynamically cold with little small dust production and the observed halo is produced in the inner disk and driven outward by radiation pressure.

### 5.2. Source of the Dust Halo

Small grains must be abundant to scatter enough light to account for our detection because of the face-on orientation of the disk (see radiative transfer modeling discussion in Wolff et al. 2023). Such grains have limited lifetimes against destruction or loss (e.g., due to radiation pressure or Poynting–Robertson drag) and must either be continuously replenished or retained on highly elliptical orbits (Müller et al. 2010). These observations show a broadly extended population of such small dust grains from at least 80–210 au. The wide planetesimal belt (80–200 au) is one potential source, though some dynamical stirring is likely required to produce this volume of dust. A planetary perturber could be responsible, but is not detected in our images. Alternatively, the stirring may have happened in the protoplanetary phase, i.e., the belts were born stirred (e.g., Wyatt et al. 2007).

While much of the dust is likely produced in collisions in the cold, outer belt, some portion of the dust could also be produced closer to the central star and transported outward by radiation pressure. Recent MIRI images of the disk (K. Su et al. 2024, in preparation) show a dust disk extending from the outer, cold belt into at least 5 au with a dip in the surface-brightness slope between 40 and 75 au. A component that large is no longer firmly associated with the ice line and might indicate processes such as a high rate of inward scattering of comets by a series of planets (Raymond & Bonsor 2014), or the operation of a mechanism analogous to that proposed by Faramaz et al. (2017) based on mean motion resonance effects in the main debris disk driven by a planet outside it. However, for the case of Vega, K. Su et al. (2024, in preparation) find that the most likely source for the distribution of dust detected interior to the broad outer belt is consistent with grains being dragged inward by the Poynting–Robertson effect. The authors are also able to rule out shepherding planets with a mass larger than $6\,M_\oplus$.

There is also a population of dust in the innermost regions of the system. The HOSTS survey for exozodiacal dust measures $33.2 \pm 7.5$ zodis of material using a conservative aperture of $0\rlap{.}{''}66$ or $\lesssim 5$ au (Ertel et al. 2020). This zodi level is low compared to some other A-type stars and cold belt hosts in their sample, which could indicate the influence of planets either to clear material locally or prevent the inward migration of dust. Regardless, this significant population of dust emitting at $\sim 10\,\mu m$ in the inner regions could serve as a reservoir for small particles being driven outward by radiation pressure. Pinpointing the source of this small dust population requires a more complete map of the inner-disk structure.

### 5.3. Comparison with Other Scattered-light Halos

We now discuss the Vega system in a wider context. Is the need for sub-blowout grains unique to the Vega scattered-light halo? Thebault et al. (2023) list 37 stars with scattered-light halos, and we base the discussion largely on this list.

There are no cases in this list with fractional luminosities, $f = L_{\rm debris}/L_{\rm star}$, less than $10^{-4}$. This compares with the value for Vega of $f \sim 3.1 \times 10^{-5}$ (Cotten & Song 2016). However, this result is less significant than it appears. Targets for observations of scattered-light halos have been selected on the basis of inferred large dust mass, closely related to $f$ (e.g., Ardila et al. 2004; Kalas et al. 2006). Schneider et al. (2014) state explicitly that their sample was selected to have fractional luminosities $>1 \times 10^{-4}$. The listing of detected halos in Thebault et al. (2023) is therefore biased against those in debris systems with smaller fractional luminosities.

One might expect a strong bias toward detections of halos around stars at high inclination, for two reasons. First, the higher surface brightness and large departures from circular symmetry of edge-on halos make them easier to identify. Second, scattering by grains above the blowout limits (1–5 $\mu m$ for most of the relevant stars) is of low efficiency at large angles, which are required for detection of low-inclination examples. However, 10 of the 37 listed are at inclinations $<60°$, whereas from simple geometric arguments one would expect 50% or 18 if there were no bias against small inclinations. To isolate the need for large-angle scattering but where tiny grains are not expected, we consider the cases with stars sufficiently luminous for radiation blowout of small grains. There are 30, of which nine are at inclinations $<60°$. If there were no observational bias toward the higher surface brightness of edge-on systems, we would expect 15. It appears that large-angle scattering in systems where small grains would be expected to be removed is a common behavior.

A few disks are at sufficiently low inclinations and have good enough images that the presence of small grains can be tested from the observed scattering at the phases around the disk. For example, for HD 107146, Ardila et al. (2004) say "the mean color is consistent with the presence of grains smaller than the radiation pressure limiting size. A more detailed analysis (with a more realistic dust model) is necessary to confirm this result, although a similar situation has been found for the debris disk around HD 141569." For HD 202628, Schneider et al. (2016) report that the ring (at an inclination of 54.2°) has "(nearly) isotropic scattering." Polarimetric study of HD 181327 by Milli et al. (2024) found that large numbers of grains smaller than 0.3 $\mu$m and far below the blowout size of a few micrometers were needed to fit the measurements.

In summary, our detection of a scattered-light halo around Vega is generally consistent with the behavior of other stars with debris disk halos. It appears that many of these halos require the presence of large numbers of tiny grains well below the blowout limits (see also Ballering et al. 2016), to provide efficient large-angle scattering. The creation of large populations of these tiny grains is the outstanding question posed by the large number of observed halos. This is a complex issue and may have different explanations for different ages and types of stars.

For example, for HD 107146, 181327, and 202628 (of spectral types F6V–G2V), a possibility for creating this grain population is that the collisional cascades among the larger particles inevitably create a broad range of particle sizes. A subset of the particles is likely to have $\beta < 0.5$ and still provide sufficient scattering cross section to be detected. The population will be winnowed by radiation pressure force to eject all the grains with $\beta > 0.5$. The loss mechanism for the remaining grains is not clear. If it is Poynting–Robertson drag, they may have long lifetimes (Gustafson 1994),

$$\tau_{P-R} \approx \frac{400}{\beta}\left(\frac{M_\odot}{M_*}\right)\left(\frac{r_0}{\mathrm{au}}\right)^2 \mathrm{yr},$$

where $\tau_{P-R}$ is the time to spiral into the star, and $r_0$ is the starting distance from the star. Thus, even if the production of such tiny grains is very inefficient, they can have very long lifetimes, allowing the accumulation of the significant reservoirs needed for the scattered-light halos.

Thebault & Kral (2019) have suggested that for disks with high fractional luminosities, the generation of very small grains may be adequate to explain their presence in large numbers, even as they are blown out of the systems. HD 141569 presents a different possibility: it is sufficiently young that residual gas may help retain small grains. However, neither of these explanations appear to apply to Vega or Fomalhaut, given their ages and much larger luminosities and the concomitant lack of small values of $\beta$ even for very small grains. One possibility is a transient event such as a major planetesimal collision or a series of collisions that are enhancing the small grain population temporarily (e.g., Su et al. 2005).

Further explorations of these possibilities for the generation of the needed grain populations are beyond the scope of this paper.

## 6. Conclusion

The HST coronagraphic imaging of the Vega system reported here demonstrates the following:

1. Vega appears to have a very extended halo in scattered light detected from 10″.5 to 27″ (80–208 au). The high level of scattering in a face-on disk indicates a significant population of small (<3 $\mu$m) grains, below the predicted blowout size.
2. Although the shape of this halo is only tentatively established due to challenges with the HST reference stars, to first order it appears to follow the large extended halo seen in thermal emission throughout the infrared, from 24 $\mu$m (Su et al. 2006; K. Su et al. 2024, in preparation) through 500 $\mu$m (Sibthorpe et al. 2010).
3. Although it has been thought that the Vega and Fomalhaut debris systems were twins (Su et al. 2013), in scattered light they are very different. In Fomalhaut the light is confined to the narrow outer debris belt, while Vega has a far more extended distribution.
4. This difference in debris disk structure suggests a difference in the architecture of the planetary systems around the stars: It is believed that the Fomalhaut ring is shepherded by planet(s), but such planets may be absent or have substantially different properties in the outer region of the Vega system.
5. This dust population may arise due to a very high rate of collisions in the outer belt of the system, but the details of its origin are not certain.

## Acknowledgments

The data presented in this paper were obtained from the Mikulski Archive for Space Telescopes (MAST) at the Space Telescope Science Institute. The specific observations analyzed can be accessed via 10.17909/wvcd-5w86. STScI is operated by the Association of Universities for Research in Astronomy, Inc., under NASA contract NAS526555. Support to MAST for these data is provided by the NASA Office of Space Science via grant No. NAG57584 and by other grants and contracts. L.M. acknowledges funding from the Irish Research Council (IRC) under grant No. IRCLA-2022-3788. Finally, we wish to thank Glenn Schneider for his assistance in planning this program and for his many contributions to disk studies with HST/STIS coronagraphy.

*Facility:* HST(STIS).

*Software:* Astropy (Astropy Collaboration et al. 2013, 2018, 2022).

## Appendix A
## A Deep Dive into the Systematic Uncertainties

Here we examine the impacts of several sources of systematic error in the PSF subtraction process described in Section 4. In summary, we find that (i) inter-orbit breathing and PSF centering errors are not significant, with <1% variations seen across the 25 individual Vega orbits; (ii) the relative scaling factor between Vega and $\alpha$ Cyg is challenging to decouple from color-induced effects inside of ~5″ but does not greatly impact the disk morphology; and (iii) a slight color mismatch between Vega and the PSF reference star $\alpha$ Cyg causes notable residuals at small separations, and could have some impact at the location of the detected disk (10″.5–30″), though this is difficult to quantify.

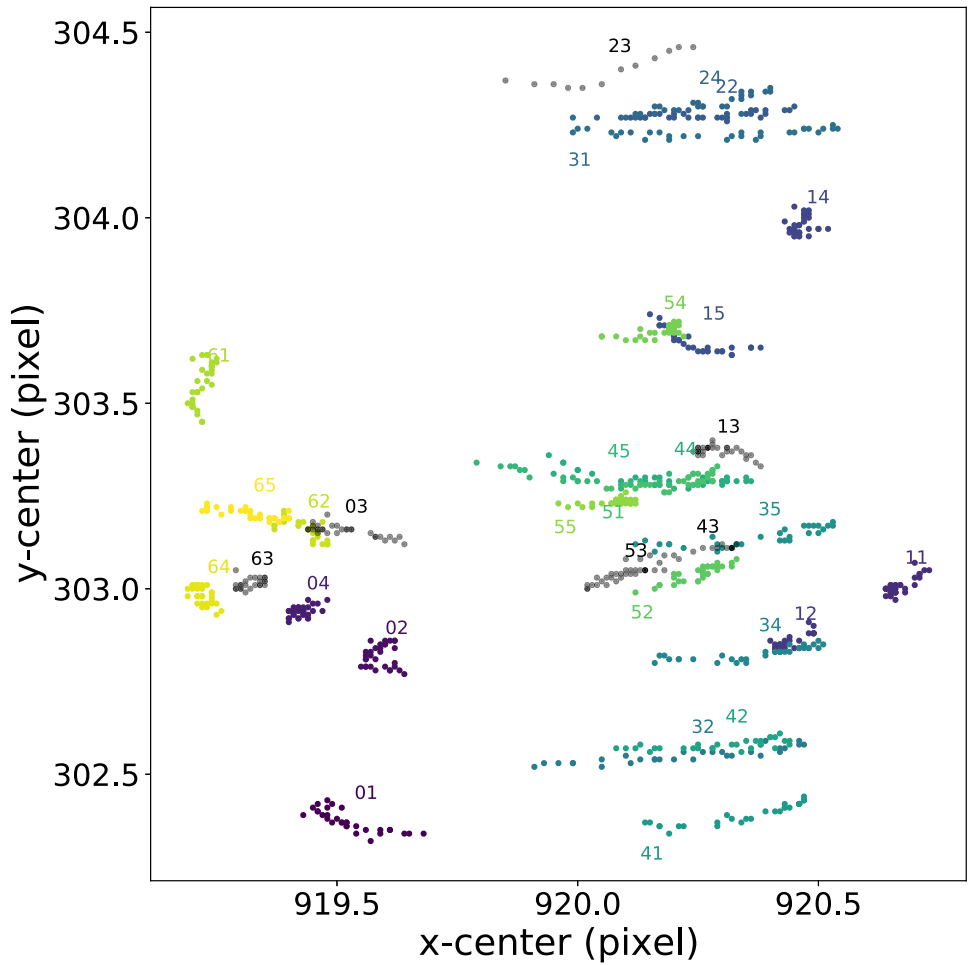

**Figure 6.** The computed stellar positions for each frame of the 25 Vega orbits and six reference orbits obtained in this program. The center positions have been color-coded based on the orbit (see Table 1) and have corresponding orbit labels. Within an orbit the center positions are very stable and remain stable over the full 11 months of observations.

*A.1. PSF Centering, Breathing, and Focusing*

The STIS coronagraph is simple, with a single mask and a Lyot stop and several preset target coronagraph positions behind the mask. Exact positioning of the star behind the PSF mask is typically done by fitting to the diffraction spikes, either by hand or algorithmically. Given the large data volume, we opted to determine the stellar position using the RadonCenter code (Ren et al. 2019a), which performs a Radon transform on the diffraction spikes and has been extensively tested on STIS coronagraphic data. We have verified the results of this algorithm to the centers derived from a "by-eye" fit to the diffraction spike locations. The computed centers for each orbit are shown in Figure 6. In general, the stellar positions are very stable over the course of an orbit, and remain stable over the program with a total offset of <2 pixels in the *x*-direction and <2.5 pixels in the *y*-direction observed. In any individual orbit, the stellar center is displaced by only 0.1–0.2 pixels.

Next we discuss another systematic uncertainty that can lead to "halo"-type PSF subtraction artifacts: OTA breathing (secondary mirror motion caused by thermal variations). Unfortunately, no model of the HST PSF at such wide separations exists for testing the effects of breathing and/or focus changes over time. However, these observations were taken in sets of contiguous orbits to limit focus/breathing changes, and very little variation is observed. As a color-independent test of breathing effects, we show the ratio of each Vega orbit relative to the median in Figure 7. Radial profiles were computed for the reduced but non-PSF-subtracted Vega observations where frames taken within each individual orbit were median combined. Inside of ∼34″, the orbits all agree well, with variations of <1%. Note that these orbit-to-orbit variations include both effects from breathing and any impacts due to the subpixel recentering performed before the subtraction.

*A.2. Relative Scaling between Vega and α Cyg*

The relative flux scaling between the science and reference targets represents another source of systematic uncertainty. The strong saturation pattern and impacts of color at small separations complicates this measurement, as does the known variability of α Cyg. Here we compare the individual orbits of Vega and α Cyg to confine the range of reasonable scale factors and discuss the impact on the final PSF-subtracted image. Figure 8 shows the radial profiles for all Vega (blue) and α Cyg orbits using various scale factors. In each case, we compute a percent difference from the median α Cyg orbit radial profile to highlight the differences. The top panel uses a scale factor of 2.99 for all Deneb orbits. This value reflects the ratio of the mean counts per frame for the full science image stack (Vega) and reference image stack ($\alpha$ Cyg). This is in line with what is expected for the difference in the count rate between Vega and $\alpha$ Cyg based on their respective magnitudes. This panel highlights the temporal variability of $\alpha$ Cyg. In the middle panel, we correct for the variability. We employ scale factors of 2.97, 2.93, 2.95, 2.95, and 2.99 for orbits 03, 13, 43, 53, and 63, respectively. While the absolute flux of the $\alpha$ Cyg orbits vary, the variability-corrected radial profiles remain very stable out to 30″. Note that there is a clear difference in the curvature between the Vega and $\alpha$ Cyg profiles. The normalized Vega orbits exhibit a peak at 4″ (likely a color-induced effect) and a broader peak from 10″ to 30″ (the source of the residual scattered light). The bottom panel inflates the scale factors used in the middle panel by 1%, which produces oversubtraction inside of 10″ and in the outer regions.

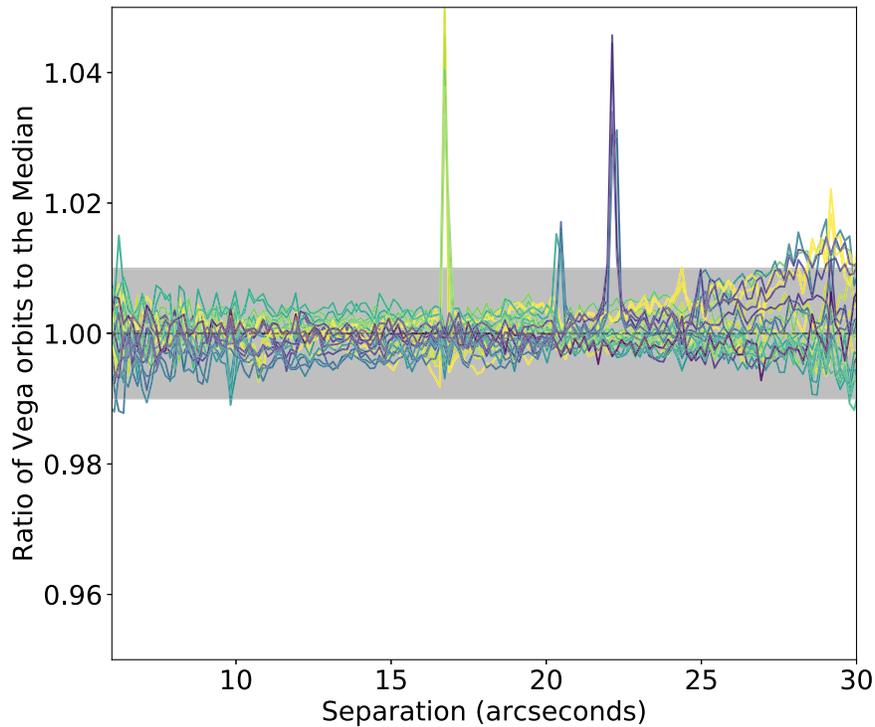

**Figure 7.** Here we show the radial profile for each of the 25 median-averaged Vega orbits divided by the median radial profile. This provides a color-independent estimation of the impacts of breathing on these observations. The telescope PSF appears stable over the ∼year of observations. Inside of 30″, variations between orbits are <1%, with no clear curvature that could mimic a scattered-light dust halo.

The middle and bottom panels of Figure 8 represent our upper and lower bounds on the scale factor for $\alpha$ Cyg and correspond to a 1% systematic uncertainty in the scale factor. The resultant images are presented in Figure 9. The bounds on the feasible scale factors were determined via trial and error. Too low a scale factor results in undersubtraction in the outer regions of the image (exterior to the observed signal) inconsistent with the difference in magnitude between Vega and $\alpha$ Cyg. Too large a scale factor resulted in oversubtraction both in the outer regions and in the inner regions (<10″) after correcting for color-induced effects. While the Vega/$\alpha$ Cyg scale factor used does impact the total flux of the Vega disk signal, it does not have a strong impact on the resultant slope of the radial profile, or on the S/N of our detection (see Appendix B). For the final image reduction presented in Section 4, we choose an intermediate 0.5% inflation of our baseline scale factors and discuss the impacts of scale factor on the total flux and derived scattering optical depth and 90° scattering albedo.

### A.3. Impacts of Color on the PSF

The dominant systematic noise source in the PSF-subtracted reduction is expected to be the slight color mismatch between Vega and our principal PSF reference source, $\alpha$ Cyg, which is discussed extensively above. It is recommended that the broadband *UVBRI* color differences be less then 0.08 magnitudes between the target and reference star when observing with the STIS coronagraph. In our case, the $\Delta R - I = 0.07$ meets this criterion while $\Delta B - V = 0.09$ is slightly over and both $\Delta V - R = 0.15$ and $\Delta U - B = 0.23$ are not ideal. This color mismatch, most notable at shorter wavelengths, was unavoidable, as there were no other reference stars available that were sufficiently bright (to achieve the required S/N within an orbit and to limit CTE degradation; Debes et al. 2019) and without bright companions in the FOV (see also Section 3). Additionally, while the wider wedge position was necessary to limit saturation around these bright, nearby stars, there are no archival color-matched STIS/WEDGEB2.8 observations available. Only one other source has been imaged using the WEDGEB2.8 position: GD 153 (a $V_{mag} = 13.3$ white dwarf, PID 7151) with very short exposures designed to verify STIS aperture updates.

In the absence of a perfect color and magnitude match for Vega, we examine the impact of stellar colors on the PSF shape using archival HST/STIS observations at wide separations. Grady et al. (2003) examine the impact of an increasing $\Delta B - V$ colors on the STIS coronagraphic profiles, and find that this produces a pattern of positive and negative rings dependent on the direction of the color mismatch. To further investigate this effect, we examine two stars in their sample, HD 88981 (Am spectral type, PID 9241) and HD 141653 (A2

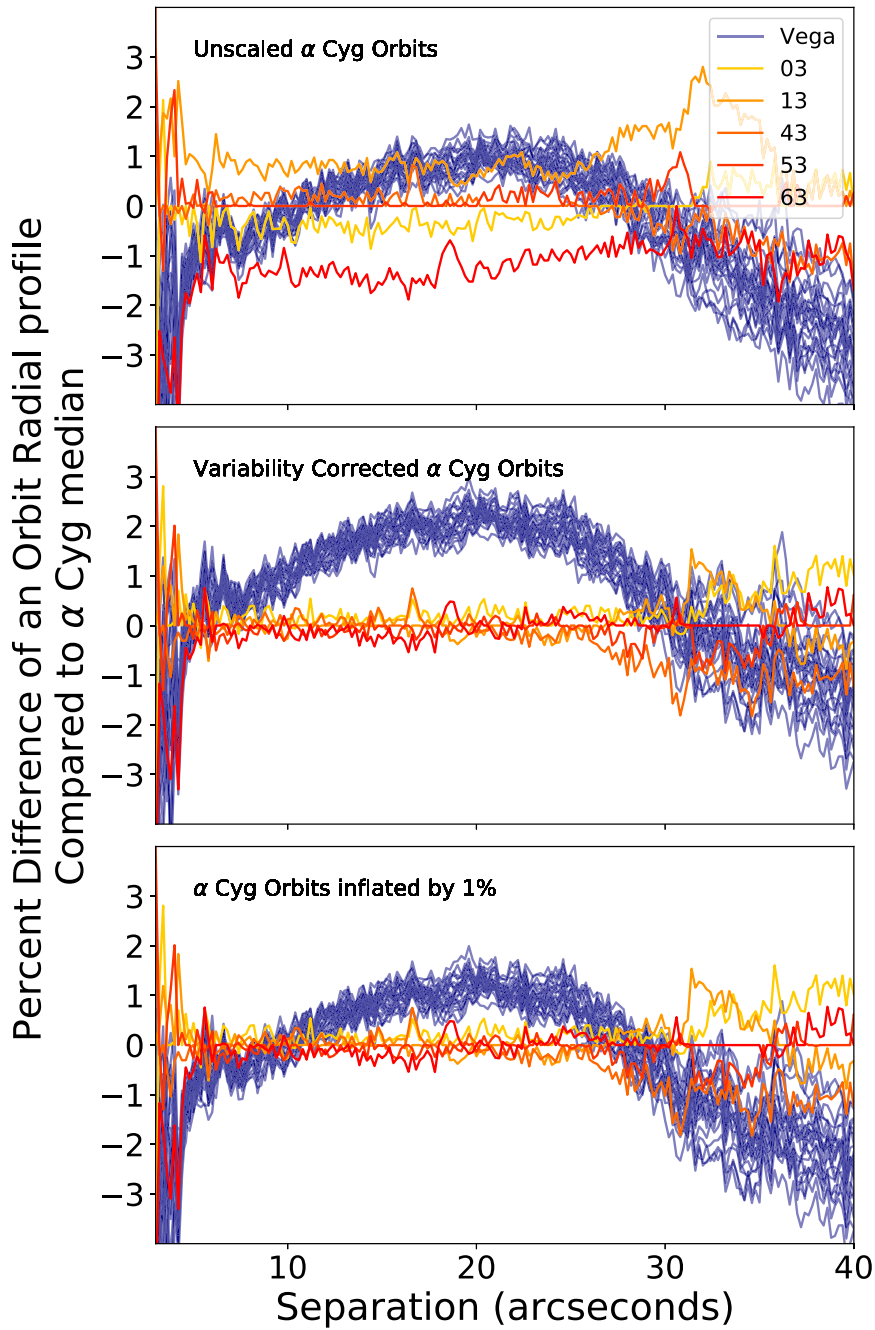

**Figure 8.** Radial profiles for each of the 32 orbits in this program prior to PSF subtraction. Vega orbits are displayed in blue and α Cyg orbits are shown in red. We display a percent difference from the median α Cyg orbit to highlight the features. Color-mismatch effects are apparent inside of ∼5″, while a broad peak in the Vega orbits is visible out to 30″. The top panel uses a uniform scale factor for the α Cyg orbits to illustrate the variability. The middle panel corrects for the variability by normalizing to orbit 63, while the bottom panel inflates these scale factors by 1%.

IV spectral type, PID 8419), with a $\Delta B - V = 0.15$. Like Vega and α Cyg, these are both A-type stars but with a greater color mismatch, and they provide a good test for the magnitude of the color-induced "corona" effect. We perform a simple subtraction, and find results that agree with those discussed in Grady et al. (2003): a negative ring near the coronagraph inner working angle, and positive residuals from ∼6″ with a power-law slope of −3 out to ∼10″ where the residuals level off, likely as the photon noise floor has been reached (see Figure 10). The scattered-light signal observed around Vega is outside of 10″.5, with a shallower slope, and is difficult to directly compare to the HD 88981 – HD 141653 reduction. However, the rapid drop in the residual signal with increasing radius approaching 10″ is encouraging.

In a further attempt to quantify color-induced corona effects at wide separations, we conducted a STIS archival search for a Vega/α Cyg color match. Observations for β Pic and a PSF reference α Pic obtained at various wedge positions (PID 16992) have similar $\Delta B - V = 0.06$ and $\Delta V - I = 0.08$ colors. While analysis of this data set is still ongoing, the authors find the lack of a residual color halo down to a few thousands of millijanskys per square arcsecond past 20″ (Avsar et al. 2024).

Our most reliable color test at a wide wedge position is the single orbit of ζ Aql data obtained during orbit 23 of our program. This was the second best PSF reference star for our Vega observations and fits the *UBVRI* color match criterion, but is 3 magnitudes fainter than Vega and has a stellar companion. A classical PSF subtraction using ζ Aql is described in detail above. While it is the best constraint on color mismatch effects at wide separations (>10″), the lower integration time and impacts on the saturation caused by the stellar companion limit its effectiveness.

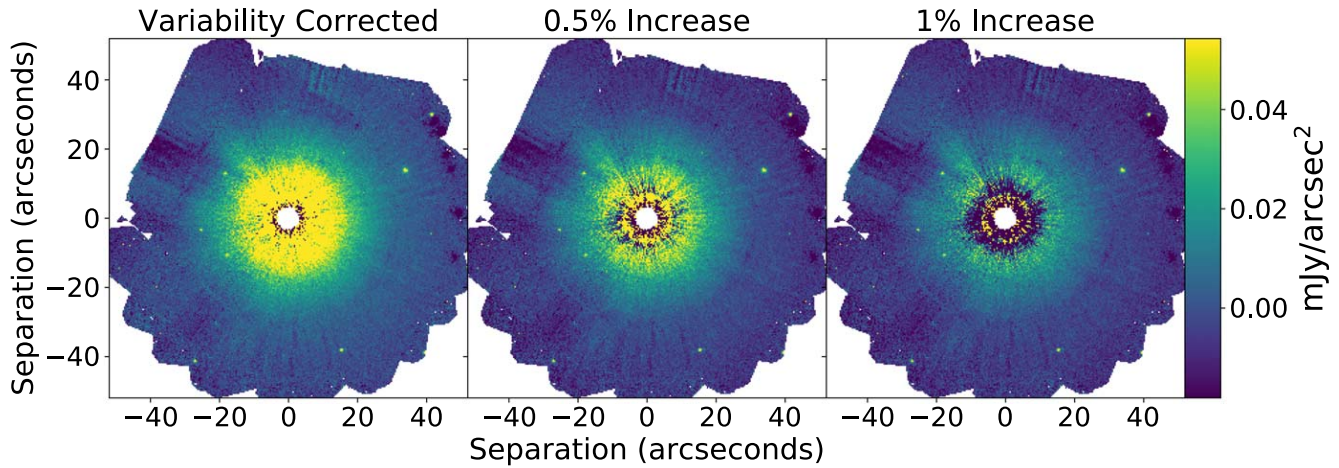

**Figure 9.** The impact of the relative scaling between the α Cyg and Vega orbits on the final reduction. The left panel shows the α Cyg − Vega reduction using the variability-corrected α Cyg orbit scale factors (equivalent to the middle panel of Figure 8). The right panel shows the α Cyg − Vega reduction assuming a 1% increase in the α Cyg scale factors, showing a strong oversubtraction in the central region. The middle panel represents a 0.5% increase to the variability-corrected α Cyg scale factors and is equivalent to the right panel of Figure 2. In all cases the disk is visible outside of ∼10″, though the integrated flux is impacted.

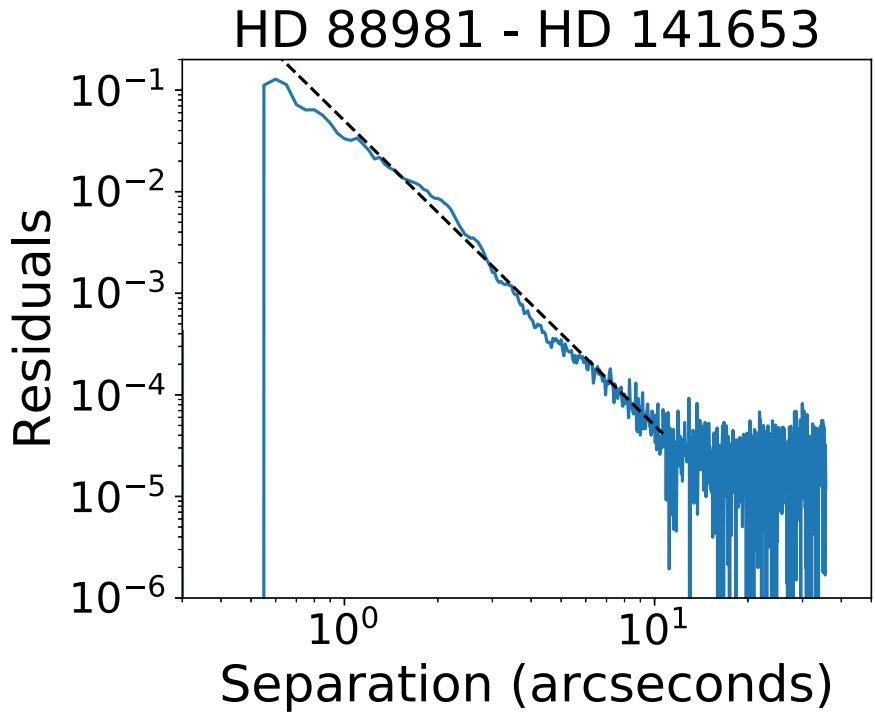

**Figure 10.** The radial profile of a quick image subtraction of HD 88981 − HD 141653 shown on a log–log scale. For each target, all images obtained at the STIS WEDGEA1.0 position were median combined, normalized to the peak, and centered using the diffraction spikes prior to subtraction. The observations show the close-in features characteristic of a color-mismatched PSF: a strong negative ring close to the inner working angle, followed by a positive "corona" with a negative power-law slope of ∼3 (black dashed line). Outside of 10″, no evidence for color-induced PSF subtraction artifacts exists, however this may be a sensitivity limitation.

The impacts of the color mismatch between Vega and α Cyg remains our dominant source of uncertainty. While no evidence for color-induced coronas outside of 10″ exists in the literature, this is a unique program in terms of the number of orbits, the degree of saturation near the central star, and the achieved sensitivity.

## Appendix B
## Noise Contributions from Random Error

In the section above, the systematic sources of uncertainty in the data set were explored. Here we focus on the known sources of heteroscedastic measurement errors, dominated by contributions from the photon noise and detector noise in both the science and reference images. A detailed assessment of the uncertainty in the reductions is supplied via a pixel coverage map, a noise map derived from the shot noise in the images, a S/N map, and finally a median absolute deviation (MAD) contrast curve.

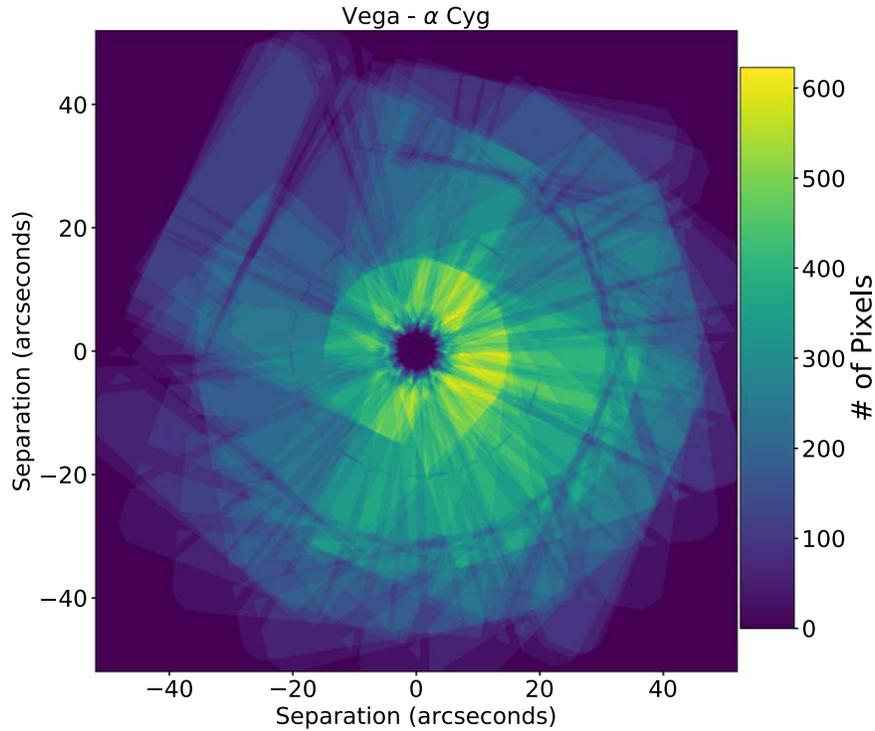

**Figure 11.** Cumulative pixel counts for the Vega − α Cyg reduction. There is generous coverage inside of 30″.

As a demonstration of the FOV coverage these 32 orbits provide, we generate cumulative pixel maps for the α Cyg reduction in Figure 11. For each unique orientation angle, the masks for the science and reference images are multiplied, and only pixels with a value of 1 are used in the final reduction (i.e., only pixels unmasked in both the science and reference images for each orientation are used). With 25 unique orientations, each with ∼31 individual exposures (see Table 1), we have ample coverage across the FOV, and particularly in the 10.″5–30″ region of regard.

Here we estimate noise and S/N maps from the data set. While the uncertainty in the individual frames is dominated by photon-counting noise, the PSF subtraction process removes much of this noise component. Instead, we follow the method for noise estimation described in Ren et al. (2019b). We begin with the set of 25 unique Vega − α Cyg PSF-subtracted images for each science orbit, oriented to the detector coordinates. To estimate the typical noise map per frame, we compute the standard deviation for each pixel. This accounts for any temporal variations in the PSF-subtracted data set while also capturing the typical photon noise contribution of the disk signal in an individual PSF-subtracted image. This standard deviation map is then replicated and rotated to match the 25 science orientations that go into the final image, and the uncertainties are added in quadrature. This results in the map shown in Figure 12. Corresponding S/N maps are shown in the right panel of Figure 2.

The disk signal is detected at an S/N of ∼2 per pixel with an S/N > 1 per pixel demonstrated from a separation of 10.″5–30″. The larger uncertainties in the center of the image make determination of an inner radius difficult, and this is left for future work.

For completeness, we perform a similar uncertainty analysis for the ζ Aql − α Cyg subtraction, shown in Figure 13. In this case, there is only a single ζ Aql orbit, so the noise map per frame is given by $\sqrt{S_{\zeta \mathrm{Aql} - \alpha \mathrm{Cyg}}}$. No extended signal is detected at sufficient S/N. Note that, unlike the Vega − α Cyg subtraction, no point sources are present in the final reduced image because the PAs of the individual ζ Aql images have been artificially changed and the stationary point sources removed via the median combination, except in the outermost regions where there is no overlap in azimuthal coverage between orbits.

Finally, we provide contrast curves for both the Vega − α Cyg PSF-subtracted image in Figure 14. We compute the median absolute deviation (MAD = median($|x_i - \tilde{x}|$) for all pixels $x_i$ in an annulus) as a measure of the achieved contrast (e.g., Schneider et al. 2014) and find a minimum surface-brightness sensitivity of $4 \times 10^{-3}$ mJy arcsec$^{-2}$ outside of ∼30″ where the background noise limit is reached. The MAD is relatively robust to outliers, especially for smaller data sets and in the presence of extended disk signal; it is therefore the better estimator for our case (see, e.g., Schneider et al. 2014). For larger samples, the MAD contrast can be related to point-source optimized contrast metrics via 1.48 MAD = 1σ. We determined the value of the MAD in 2 pixel wide annuli (∼1 resolution element), centered on Vega. The interpretation of contrast curves in regions that may contain disk signal is challenging, with no clear standard in the literature. Therefore, we do not attempt to translate these contrast curves into planet sensitivity limits. NIRCam observations obtained as part of a GTO program (PID 1193; PI: Beichman) will provide deeper planet detection limits at these wide separations.

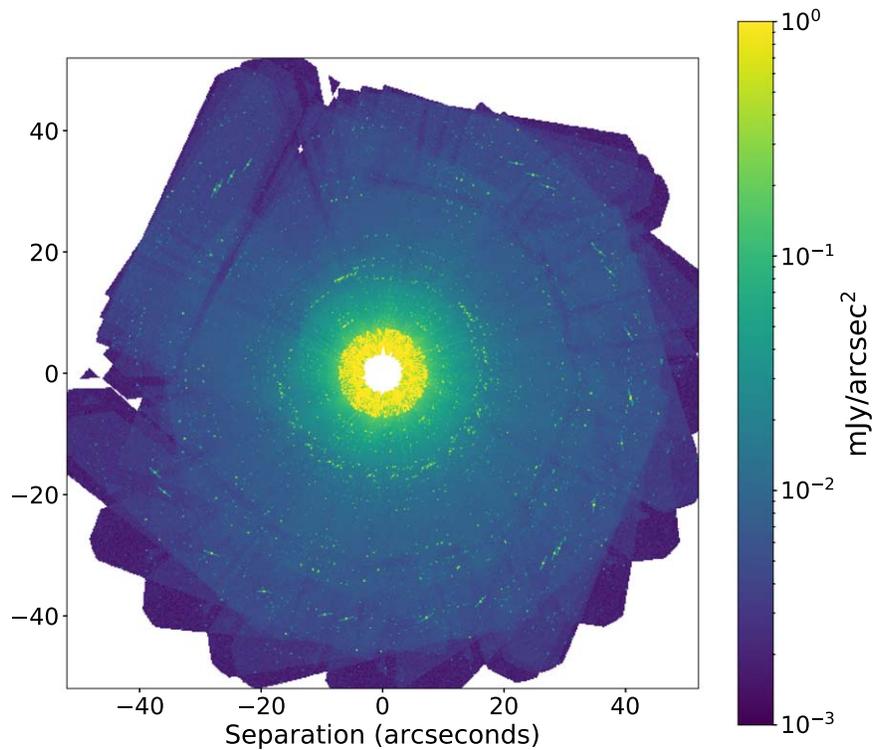

**Figure 12.** The uncertainty map derived for the $\alpha$ Cyg PSF-subtracted Vega image. The uncertainty was computed by measuring the pixel-wise standard deviation for the full stack of de-rotated PSF-subtracted images to estimate temporal variations. This standard deviation map was then replicated and rotated to match the azimuthal coverage of the Vega data set. Multiple rotations coupled with several point sources in the Vega FOV lead to azimuthally symmetric artifacts.

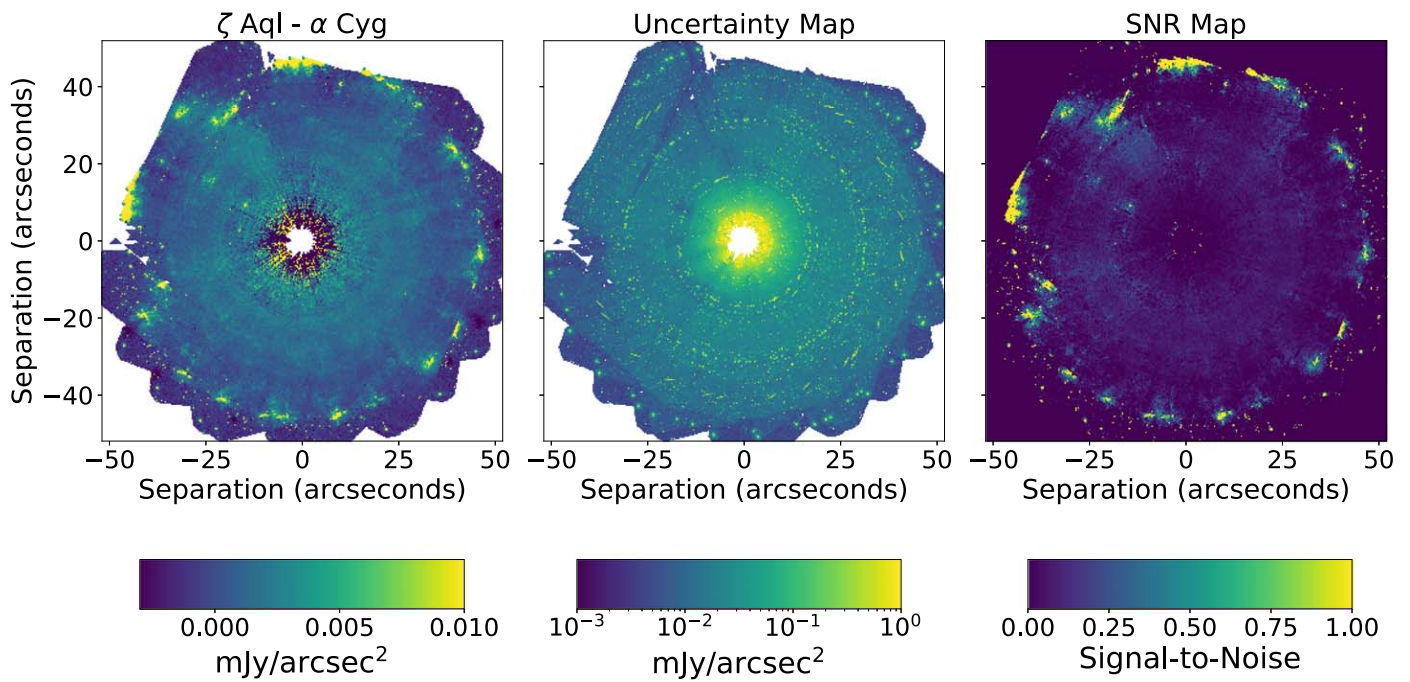

**Figure 13.** Here we present the final cRDI PSF-subtracted image using $\zeta$ Aql as the science target and $\alpha$ Cyg as the reference PSF. Left: the cRDI-subtracted image shown on a linear scale. Middle: the uncertainty map generated using the same technique as shown in Figure 12 shown on a logarithmic scale. Right: the S/N map for this reduction. In all cases, the images are rotated north up. The PSF-subtracted image shows strong negative residuals inside of 10″, consistent with the minor color mismatch between $\zeta$ Aql $\alpha$ Cyg. The S/N map shows no detection of extended signal outside of this region.

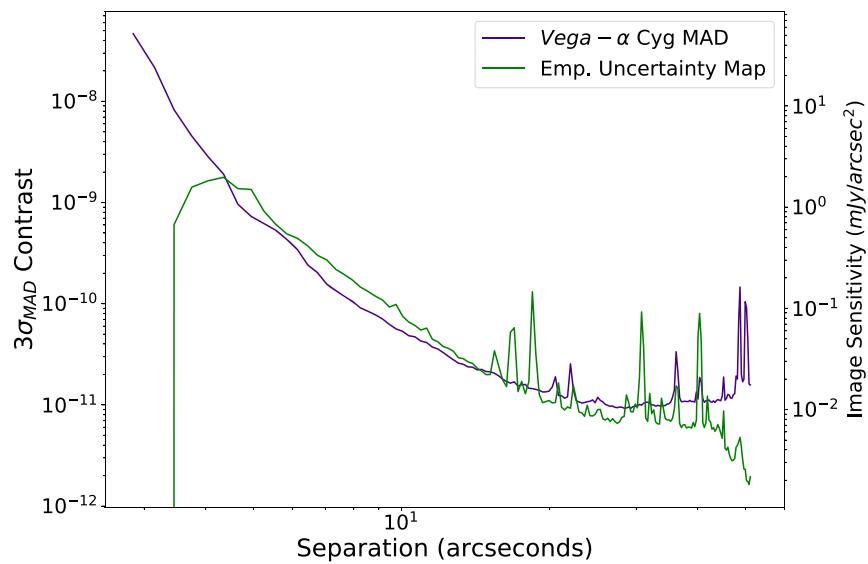

**Figure 14.** The 3$\sigma$ median absolute deviation (MAD) contrast for the Vega − $\alpha$ Cyg reduction (purple). The 32 orbits of the science target, Vega, enable among the lowest achieved MAD contrasts to date. Image sensitivities are shown on the left. For comparison, we also include the image sensitivities associated with the radial profile of the empirically estimated uncertainty map for the Vega − $\alpha$ Cyg reduction shown in Figure 12 (green).

## ORCID iDs

Schuyler G. Wolff ⓘ https://orcid.org/0000-0002-9977-8255
András Gáspár ⓘ https://orcid.org/0000-0001-8612-3236
George H. Rieke ⓘ https://orcid.org/0000-0003-2303-6519
Jarron M. Leisenring ⓘ https://orcid.org/0000-0002-0834-6140
Kate Su ⓘ https://orcid.org/0000-0002-3532-5580
David J. Wilner ⓘ https://orcid.org/0000-0003-1526-7587
Luca Matrà ⓘ https://orcid.org/0000-0003-4705-3188
Marie Ygouf ⓘ https://orcid.org/0000-0001-7591-2731
Nicholas P. Ballering ⓘ https://orcid.org/0000-0002-4276-3730